\documentclass[aps,prx,reprint, amsmath, amssymb]{revtex4-1}

\pdfoutput=1
\usepackage[utf8]{inputenc}
\usepackage{bm}
\usepackage[english]{babel}
\usepackage[T1]{fontenc}
\usepackage{mathrsfs}
\usepackage[retainorgcmds]{IEEEtrantools}
\usepackage{amsmath}
\usepackage{amssymb}
\usepackage{color}
\usepackage{amsfonts}
\usepackage{times,txfonts}
\usepackage{nicefrac}
\usepackage[colorlinks=true,linkcolor=blue,urlcolor=blue,citecolor=blue,pdfusetitle]{hyperref}

\renewcommand{\r}{\hat{\rho}}
\newcommand{\dd}{\mathrm{d}}
\newcommand{\ii}{\ensuremath{\mathrm{i}}}
\newcommand{\ee}{\ensuremath{\mathrm{e}}}
\newcommand{\od}[2]{\ensuremath{\frac{\dd #1}{\dd #2}}}
\newcommand{\pd}[2]{\ensuremath{\frac{\partial #1}{\partial #2}}}

\usepackage{float}
\usepackage{epsfig} 
\usepackage{subfigure}
\graphicspath{{Images/}}

\usepackage{physics}
\usepackage{braket}
\usepackage{amssymb}
\usepackage[amssymb]{SIunits}

\usepackage{xcolor}

\begin{document}

\title{Extractable work in quantum electromechanics}

\author{Oisín Culhane}
\email{oculhane@tcd.ie}
\affiliation{Department of Physics, Trinity College Dublin, Dublin 2, Ireland}
\author{Mark T. Mitchison}
\email{mark.mitchison@tcd.ie}
\affiliation{Department of Physics, Trinity College Dublin, Dublin 2, Ireland}
\author{John Goold}
\email{gooldj@tcd.ie}
\affiliation{Department of Physics, Trinity College Dublin, Dublin 2, Ireland}

\begin{abstract}
Recent experiments have demonstrated the generation of coherent mechanical oscillations in a suspended carbon nanotube, which are driven by an electric current through the device above a certain voltage threshold, in close analogy with a lasing transition. We investigate this phenomenon from the perspective of work extraction, by modelling a nano-electromechanical device as a quantum flywheel or battery that converts electrical power into stored mechanical energy. We introduce a microscopic model that qualitatively matches the experimental finding, and compute the Wigner function of the quantum vibrational mode in its non-equilibrium steady-state. We characterise the threshold for self-sustained oscillations using two approaches to quantifying work deposition in non-equilibrium quantum thermodynamics: the ergotropy and the non-equilibrium free energy. We find that ergotropy serves as an order parameter for the phonon lasing transition. The framework we employ to describe work extraction is general and widely transferable to other mesoscopic quantum devices. 
\end{abstract}

\maketitle

%%%%%%%%%%%%%%%%%%%%%%%%%%%%%%%%%%%%%%%%%%%%%%%%%%%%%%%%%%%%%%%%%%%%%%%%
%
%
%
%
%%%%%%%%%%%%%%%%%%%%%%%%%%%%%%%%%%%%%%%%%%%%%%%%%%%%%%%%%%%%%%%%%%%%%%%%

At its core, thermodynamics describes the relation between heat and mechanical motion~\cite{Thompson1798,Joule1850}. Recent advances in the control and measurement of mechanical motion in the quantum regime~\cite{Barzanjeh2021,Ananyeva2021,Tebbenjohanns2021} provoke fascinating questions about the fundamental limits of heat-to-work conversion in microscopic systems~\cite{Binder2018}. To explore these questions, mechanical heat engine and refrigerator cycles have been demonstrated in proof-of-principle experiments on single trapped ions~\cite{Rossnagel2016,Lindenfels2019,Maslennikov2019} and nanomechanical oscillators~\cite{Steeneken2011, Klaers2017}. Significant experimental progress has also been made in measuring and harvesting heat energy in nanoscale electronic devices~\cite{Hartmann_2015,Halbertal_2016,Roche_2015,Pekola_2015,Forneiri_2017,Ronzani_2018,Josefsson2018,Jaliel_2019}, strongly motivated by the prospect of efficient thermoelectric energy conversion~\cite{Benenti2017}.

Nano-electromechanical systems (NEMS) are a particularly interesting platform for quantum thermodynamics, since they incorporate vibrational degrees of freedom into an electronic device ~\cite{Ares2016,ArrangoizArriola2019,OConnell2010,Rossi2018,Lee2010,Naik2006,Karwat_2018}. While such vibrations may strongly affect thermoelectric efficiency~\cite{Koch2004,Leijnse2010,Entin-Wohlman2010,Perroni2014,Perroni2016}, mechanical motion can also activate electron transport through resonant charge pumping~\cite{Leek2005,Buitelaar2008} or electron shuttling~\cite{Gorelik1998,Pistolesi2006,Wachtler_2019,Strasberg2021} mechanisms. Recently, a series of remarkable experiments have demonstrated the obverse phenomenon, in which electron transport excites coherent mechanical oscillations above a certain threshold voltage~\cite{Wen2019,Urgell2019}. The emergence of self-sustained oscillations in this context does not involve resonance between electron tunnelling and vibrations, and it can be understood as a kind of phonon lasing transition driven by non-equilibrium charge fluctuations~\cite{Blanter2004,*Blanter2004err,Clerk2005,Bennett2006}. The result is that electrical power supplied by the leads is converted into mechanical energy stored by the oscillator, which therefore behaves as an electromechanical flywheel or battery.

The storage and extraction of energy using quantum batteries is currently the focus of intense interest, due to the potential of exploiting collective quantum effects~\cite{Alicki2013, Hovhannisyan2013,Binder2015,PerarnauLlobet2015,Giorgi2015} to boost performance. Quantum batteries are typically far from equilibrium and strongly influenced by fluctuations and environmental noise~\cite{Andolina2018,Farina2019,Donato2019,Pirmoradian2019}, leading to the development of advanced control strategies to mitigate these effects~\cite{Santos2019,Gherardini2020,Quach2020,Mitchison2021a}. However, there remains a wide gap between the existing theoretical literature --- which has largely focussed on quantum-optical settings~\cite{Le2018,Ferraro2018} --- and the physics of mesoscopic electronic devices.

Here, we bridge this gap by quantifying the extractable work deposited in a self-sustained electromechanical oscillator. We first introduce a model that captures the salient aspects of the experiments reported in Refs.~\cite{Wen2019,Urgell2019}. Working in the experimentally relevant regime of slow mechanical motion relative to fast electron tunnelling, we derive a Fokker-Planck equation for the Wigner function of the oscillator, which can be efficiently solved using the quasi-adiabatic Langevin equation obtained previously by numerous authors~\cite{Mozyrsky2006,Clerk2005, Bennett2006,Metelmann2011,Blanter2004,*Blanter2004err,Blencowe2005, Usmani2007,Wachtler_2019}. We are thus able to reconstruct the quantum state of the oscillator and analyse the deposition of energy using modern tools of quantum non-equilibrium thermodynamics. We find that the states above and below threshold are starkly different in terms of their work storage potential. In particular, the ergotropy --- which bounds the work extractable by cyclic unitary operations --- behaves as an order parameter for the phonon lasing transition. Our results elucidate the predominant role of fluctuations in nano-electromechanical energy conversion, and take the first step towards a comprehensive description of the storage and extraction of electrical power from non-equilibrium quantum states of motion. We use units where the elementary charge $e$ and the reduced Planck constant $\hbar$ are equal to 1 throughout.

\begin{figure}[t]
		% Center the figure.
		\begin{center}
		% Include the eps file, scale it such that it's width equals the column width. You can also put width=8cm for example...
		\includegraphics[width=\columnwidth]{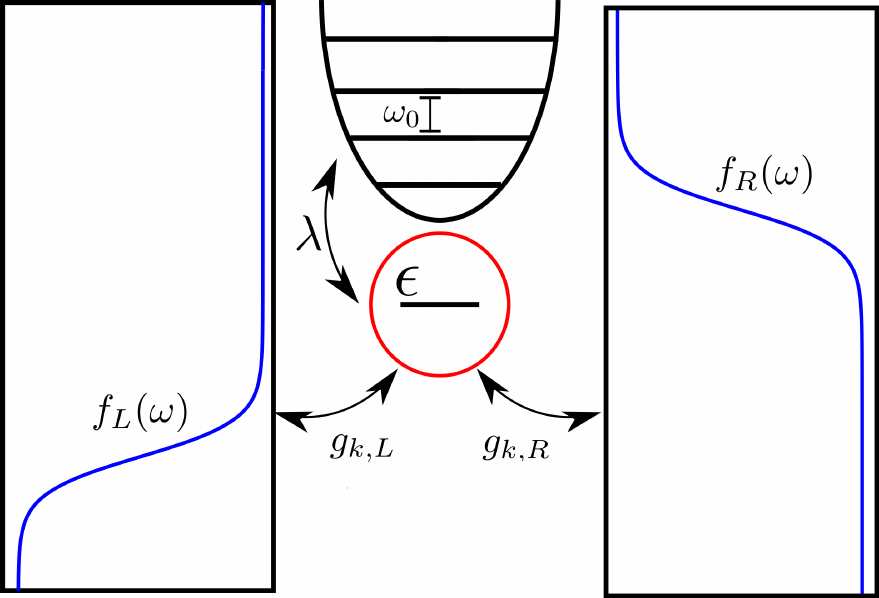}
		% Create a subtitle for the figure.
		\caption{Schematic depiction of the electromechanical system. A central quantum dot is coupled to two electrodes as well as a quantum harmonic oscillator representing the mechanical degree of freedom. The electrodes are described by Fermi-Dirac distributions $f_\alpha(\omega)$ at the same temperature but different chemical potentials. Electrons tunnelling through the quantum dot under this voltage bias excite self-sustained oscillations of the mechanical motion.\label{fig:schematic}}
		% Define the label of the figure. It's good to use 'fig:title', so you know that the label belongs to a figure.
		\label{fig:tf_plot}
		\end{center}
	\end{figure}

\textit{Model.---}To model the emergence of self-sustained oscillations, we consider the system depicted in Fig.~\ref{fig:schematic}: a single resonant electronic level sandwiched between two macroscopic leads, where the resonant level also interacts with a harmonic vibrational mode. The Hamiltonian of the model reads $\hat{H} = \hat{H}_S + \hat{H}_B + \hat{H}_T + \hat{H}_V,$ with
\begin{align}
\label{ham_quantum_dot}
    & \hat{H}_S = \epsilon \hat{c}^\dagger \hat{c}, \\
\label{ham_baths}
    & \hat{H}_B = \sum_k \left(\Omega_{kL}\hat{d}^\dagger_{kL}\hat{d}_{kL} + \Omega_{kR}\hat{d}^\dagger_{kR}\hat{d}_{kR} \right) ,\\
\label{ham_tunnelling}
    & \hat{H}_T = \sum_k \left( g_{kL}\left[\hat{c}^\dagger \hat{d}_{kL} + \hat{d}^\dagger_{kL} \hat{c}\right] + g_{kR}\left[\hat{c}^\dagger \hat{d}_{kR} + \hat{d}^\dagger_{kR} \hat{c}\right] \right),\\
\label{ham_vibrational}
   & \hat{H}_V = \frac{\hat{p}^2}{2m} + \frac{m\omega_0^2 \hat{x}^2}{2} - F\hat{n}\hat{x}.
\end{align}
Here, $\hat{H}_S$ describes a resonant level with energy $\epsilon$ and fermionic annihilation operator $\hat{c}$. Meanwhile, $\hat{H}_B$ models the leads as collections of non-interacting fermions described by annihilation operators $\hat{d}_{k\alpha}$ and dispersion relations $\Omega_{k\alpha}$, where $\alpha = L,R$ denotes the left or right lead. The tunnelling from the leads to the resonant level is described by $\hat{H}_T$, with coupling constants $g_{k\alpha}$. The effect of the leads on the central system is fully characterised by their initial temperature, $T$, chemical potentials, $\mu_\alpha$, and spectral densities (level-width functions) $\kappa_\alpha(\omega) = 2\pi\sum_k g_{k\alpha}^2\delta(\omega - \Omega_{k\alpha})$. Finally, $\hat{H}_V$ describes a vibrational mode with mass $m$, angular frequency $\omega_0$, momentum $\hat{p}$ and position $\hat{x}$. This oscillator experiences an electrostatic force proportional to the excess charge localised on the resonant level, $\hat{n} = \hat{c}^\dagger \hat{c} - N_0$, where $F$ is the force per unit charge and $N_0 = \langle \hat{c}^\dagger \hat{c}\rangle\rvert_{F=0}$ is the level occupation in the absence of electromechanical coupling. Our simplified model --- which neglects spin degrees of freedom and electronic interactions, as well as intrinsic non-linearities and damping of the vibrational motion --- suffices to qualitatively reproduce the physics of mechanical oscillations driven by electron tunnelling.
% % The final term consists of the mechanical degrees of freedom of the system. We model the vibrational degrees of freedom to be a single mode with frequency $\omega_0$, the coupling term couples the occupation of the quantum dot to the position of the oscillator. Here we include the main approximation used in this paper. We are in the quasi-adiabatic limit, the electrons on the quantum dot see a stationary value for the oscillator, the electronic degrees of freedom relax much faster than the mechanical degrees of freedom ie 
% \begin{equation}
% \label{adiabatic}
%     \Gamma \gg \omega_0,\lambda
% \end{equation}
We focus on the quasi-adiabatic limit, in which the motion of the oscillator is slow in comparison to the characteristic rate of electron tunnelling $\Gamma$ and the thermal correlation time, $\beta = 1/k_BT$, i.e.~
\begin{equation}
\label{adiabatic_condition}
    k_BT, \Gamma \gg \omega_0,|\lambda|,
\end{equation}
where $\lambda = Fx_0/2$ characterises the electromechanical interaction energy and $x_0 = (2/m\omega_0)^{-1/2}$ is the mechanical zero-point fluctuation. For example, the experimental parameters of Ref.~\cite{Wen2019} are on the order of $\Gamma \sim 10~$GHz, $k_BT \sim 1$~GHz, $\omega_0
\sim 0.1$~GHz and $\lambda\gtrsim 0.1$~GHz, for which Ineq.~\eqref{adiabatic_condition} holds. We can thus derive a Fokker-Planck equation for the Wigner function of the oscillator by adiabatic elimination of the fast electronic degrees of freedom~\cite{Gardiner}; see Appendix
%~\ref{app:adiabatic_elim}
A~~\footnote{See Supplemental Material for details of numerical calculations, derivations of the Langevin equation, occupation, fluctuation and dissipation functions, and an analysis of the occupation, fluctuation and dissipation functions.} for details. When the Wigner function of the oscillator is positive, it can be efficiently sampled by solving the Langevin equation specified below. An equivalent Langevin equation has previously been derived by several authors using path-integral methods~\cite{Mozyrsky2006,Metelmann2011}, Keldysh non-equilibrium Green functions (NEGF)~\cite{Clerk2005, Bennett2006}, or master equations within the sequential-tunnelling approximation~\cite{Blanter2004,*Blanter2004err,Blencowe2005, Usmani2007,Wachtler_2019}. The advantages of our approach are three-fold: (i)~it provides direct access to the quantum state of the oscillator, (ii)~it makes a clear analogy with the theory of fluctuating laser light developed in terms of quasi-probability distributions ~\cite{Lax1967,Haake1983,GardinerZoller}, and (iii)~it can be extended to an arbitrary interacting electronic system so long as the quasi-adiabatic condition~\eqref{adiabatic_condition} is satisfied.

In the quasi-adiabatic regime, the local electronic degrees of freedom can be assumed to quickly relax to a stationary state, leading to a mean electrostatic force that depends on the oscillator's position. Furthermore, rapid tunnelling of electrons to and from the leads induces local charge fluctuations, generating both friction and an erratic Brownian force acting on the oscillator. The resulting dynamics is described by the Langevin equation
\begin{equation}
\label{langevin_equation}
    m\ddot{x} + m\gamma(x)\dot{x} + m\omega_0^2x = F\langle \hat{n}\rangle_x + \sqrt{D(x)}\xi(t),
\end{equation}
where $F\langle \hat{n}\rangle_x$ describes the average force induced by the localised excess charge, $\hat{n}=\hat{c}^\dagger\hat{c}-N_0$, and $\xi(t)$ is a zero-mean white noise. The damping rate $\gamma(x)$ and diffusion coefficient $D(x)$ are given by
\begin{equation}
\label{diffusion_damping}
    D(x)=S_x(0),\qquad  m\gamma(x) = \frac{\text{d}S_x(\omega)}{\text{d}\omega} \bigg\rvert_{\omega=0},
\end{equation}
where $S_x(\omega)$ is the noise spectrum of the fluctuating force:
\begin{equation}
\label{noise_function}
    S_x(\omega) = F^2\int^\infty_{-\infty}\text{d}t\, e^{i\omega t} \left[\langle \hat{n}(t) \hat{n}(0)\rangle_x - \langle \hat{n}\rangle_x^2\right].
\end{equation}
Here, $\hat{n}(t)$ is a Heisenberg-picture operator evolving under the effective electronic Hamiltonian $\hat{H}_x = \hat{H}_S + \hat{H}_B + \hat{H}_T - F x\hat{n}$, in which $x$ appears as a static parameter, while $\langle \bullet \rangle_x$ denotes an average with respect to the electronic steady state reached after long-time evolution under $\hat{H}_x$. Therefore, the Langevin equation~\eqref{langevin_equation} is fully specified by replacing $\hat{x}\to x$ in Eq.~\eqref{ham_vibrational} and then solving the corresponding electronic problem to find $\langle \hat{n}\rangle_x$ and $S_x(\omega)$ at each value of~$x$.

\begin{figure*}
\begin{center}
    \hspace*{-0.5cm}
\includegraphics[width=\linewidth]{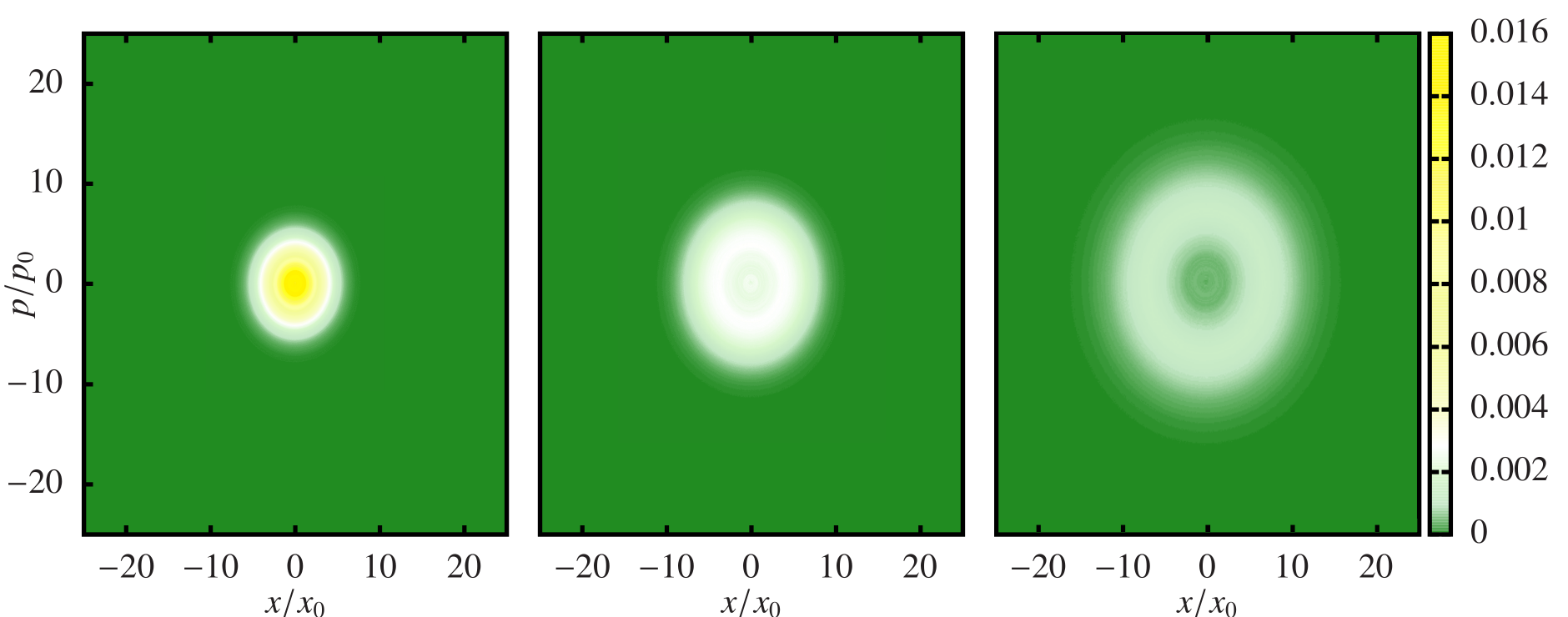}\par
\caption{Wigner functions of the oscillator at various biases. The left panel shows the Wigner function at zero applied bias showing the Wigner blob, the central panel shows the Wigner function at 6 applied bias just after the threshold voltage, the right panel shows the Wigner function at 16 bias where the voltage has reached its saturation point. Parameters: $\omega_0=0.2$,$m=1$,$\lambda=0.1$,$\omega_L=0.5$,$\omega_R=-0.5$,$\delta_L=\delta_R=1$,$\Gamma_L=\Gamma_R=2$,$\beta_L=\beta_R=0.5$, $x_0 = \left(2/m\omega_0\right)^{1/2}$, and $p_0 = \left(2m\omega_0\right)^{1/2}$.\label{fig:Wigner_function}}
\end{center}
\end{figure*} 

Self-sustained oscillations arise in the presence of negative damping rates, so that $\gamma(x) < 0$ for some $x$. In the quasi-adiabatic limit, this can only occur when the leads are out of equilibrium with each other and their spectral densities are energy-dependent~\cite{Blanter2004,*Blanter2004err,Clerk2005}. For simplicity, we assume a single temperature $T$, a finite voltage bias $V=\mu_L- \mu_R$, and Lorentzian spectral densities of the form
\begin{equation}
    \label{Lorentzian_spectral_density}
    \kappa_\alpha(\omega) = \frac{\Gamma\delta^2}{(\omega-\omega_\alpha)^2+\delta^2},
\end{equation}
which describe electronic bands centred around frequency $\omega_\alpha$ with effective bandwidth $\delta$ and overall coupling strength $\Gamma$. The electronic steady-state properties in this model are computed analytically using NEGF methods in Appendix B.

\textit{Self-sustained oscillations.---}From here on, we focus on the steady state reached by the mechanical oscillator at long times. Figure~\ref{fig:Wigner_function} shows the steady-state Wigner function obtained by sampling solutions of the Langevin equation~\eqref{langevin_equation} for three different voltages; see Appendix A
% ~\ref{app:langevin} 
for details. At zero applied voltage, the oscillator is in equilibrium with the two electrodes and the Wigner function forms a thermal blob near the origin [Fig.~\ref{fig:Wigner_function}, left panel]. This qualitative structure persists for all voltages below a certain threshold. Above this threshold, however, the damping rate $\gamma(x)$ becomes negative for small $x$, leading to amplification of the mechanical motion in direct analogy with a lasing transition. This results in an annular Wigner function [Fig.~\ref{fig:Wigner_function}, central panel], indicating the presence of population inversion. The radius of the annulus increases with $V$ until a saturation voltage is reached, beyond which any further voltage increase has no effect.

% The oscillator is given by the Hamiltonian $\hat{H}_b = \omega_0 \hat{a}^\dagger\hat{a}$. 
	\begin{figure}[b]
		% Center the figure.
		\begin{center}
		% Include the eps file, scale it such that it's width equals the column width. You can also put width=8cm for example...
% 		\hspace*{-0.6cm}
		\includegraphics[width=\columnwidth]{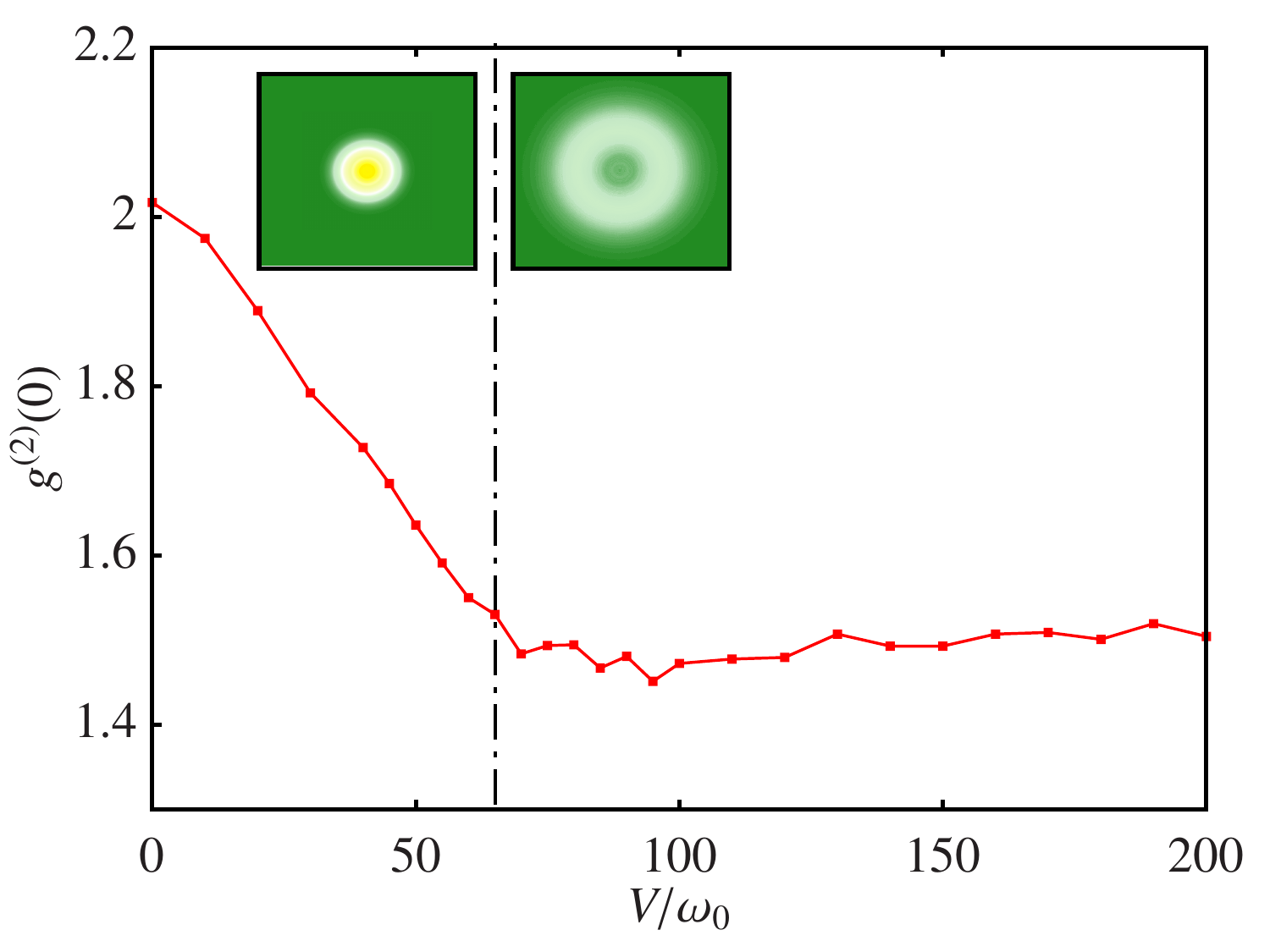}
		% Create a subtitle for the figure.
		\caption{Second-order coherence function of the oscillator as a function of voltage, for the same parameters as in Fig.~\ref{fig:Wigner_function}. \label{fig:Autocorrelation}\label{fig:tf_plot}}
		% Define the label of the figure. It's good to use 'fig:title', so you know that the label belongs to a figure.
		
		\end{center}
	\end{figure}

For further analysis we numerically transform the steady-state Wigner function into a density matrix within a truncated basis; see Appendix D 
%~\ref{app:Wigner} 
for details. Note that the Wigner functions are rotationally symmetric in phase space [see Fig.~\ref{fig:Wigner_function}], yielding density matrices that are diagonal in the energy eigenbasis of the oscillator Hamiltonian $\hat{H}_b = \omega_0 \hat{a}^\dagger\hat{a}$. It follows that the mode quadratures vanish on average, $\langle \hat{a}\rangle = 0$, in direct analogy with the quantum state of a laser field, which approximates a phase-averaged coherent state~\cite{GardinerZoller} in the absence of explicit symmetry breaking~\footnote{The relevant symmetry for an optical laser is $U(1)$ invariance associated with the photon-number superselection rule~\cite{Javanainen1996,Molmer1997,Sanders2003}. The present setup enjoys a $\mathbb{Z}_2$ symmetry corresponding to a combined parity and particle-hole transformation that exchanges the two reservoirs: $\hat{x}\to-\hat{x}$, $\hat{p}\to-\hat{p}$, $\hat{c}\to \hat{c}^\dagger$, $\hat{d}_{k\alpha}\to \hat{d}_{k\alpha}^\dagger$, and $L\leftrightarrow R$. This transformation leaves the steady state invariant for the parameters we consider, where $\mu_L+
\mu_R = \omega_L + \omega_R = 0$.}. The coherent nature of the steady state above threshold is instead conveyed through the second-order coherence function~\cite{GardinerZoller}
\begin{equation}
   g^{(2)}(0) = \frac{\langle \hat{a}^\dagger\hat{a}^\dagger\hat{a}\hat{a} \rangle}{\langle\hat{a}^\dagger\hat{a}\rangle^2},
\end{equation}
where $\hat{a} = \hat{x}/x_0 + {\rm i} \hat{p}/p_0$ is the annihilation operator for the vibrational mode. Given the diagonal steady-state density matrix $\hat{\rho}_b = \sum_n p_n\ket{n}\bra{n}$, one has $g^{(2)}(0) = \left(\sum_n n(n-1)p_n\right)/\left( \sum_n np_n \right)^2$. In Fig.~\ref{fig:Autocorrelation}, we plot $g^{(2)}(0)$ in the steady state as a function of voltage. At no applied bias, $g^{(2)}(0)\approx 2$, as expected for a thermal state. Increasing the voltage, the oscillator is pushed away from thermal equilibrium and the second-order coherence function decreases, reaching a plateau at $g^{(2)}(0)\approx 1.5$ above threshold. This indicates the sub-thermal amplitude noise associated with lasing, but is nonetheless greater than the value of $g^{(2)}(0)=1$ for an ideal coherent state, highlighting the importance of fluctuations in this nanoscale non-equilibrium system.

\textit{Extractable~work.---}The electromechanical set-up described here is an ideal platform to study work storage and extraction, central concepts in the field of quantum thermodynamics. The non-equilibrium steady state of the device gives rise to a non-thermal state in the vibrational mode with a total energy that depends on the voltage applied. The vibrational degree of freedom thus serves as a battery or flywheel that stores energy deposited by the electronics. However, the division of this energy into useful work and waste --- as per the first law of thermodynamics --- is far from obvious, since some of the deposited energy merely contributes to increasing the entropy of the oscillator, i.e.~heating.

We consider two different approaches to quantifying the deposited work. The first is the ergotropy, which gives the maximum work that can extracted from a quantum battery by a cyclic unitary transformation~\cite{Allahverdyan_2004}. Let the state of the battery be given by $\hat{\rho}_b=\sum_{k}r_k|r_k\rangle\langle r_k|$, with $r_{k}\ge r_{k+1}$, while the Hamiltonian may be written in the energy eigenbasis as $\hat{H}_b=\sum_{k}\epsilon_{k}|\epsilon_{k}\rangle\langle \epsilon_{k}|$, with $\epsilon_{k}\le\epsilon_{k+1}$ the energies. The battery state $\hat{\rho}_b$ is called passive with respect to $\hat{H}_b$ if its energy cannot be decreased by a cyclical variation of the Hamiltonian parameters over a fixed time interval $t\in[t_1,t_2]$, such that $\hat{H}(t_1)=\hat{H}(t_2) = \hat{H}_b$. Mathematically, the conditions for passivity are that $[\hat{\rho}_b,\hat{H}_b]=0$ and $r_n\ge r_m$ whenever $\epsilon_n<\epsilon_m$. If either of these conditions are violated then work may be extracted. By optimising over all possible protocols, one obtains an upper bound on the extractable work known as the ergotropy~\cite{Allahverdyan_2004}:
\begin{equation}
    \mathcal{W}_E = \sum_{j,k} r_j\epsilon_k \left(\lvert\braket{r_j|\epsilon_k}\rvert^2 - \delta_{jk} \right).
\end{equation}

Such unitary work extraction is rather idealised since the requisite protocol would often be challenging to implement. We thus consider a second approach to work extraction based on the non-equilibrium free energy~\cite{Donald_1987,Vedral_2002,Esposito_2011,Horodecki_2013}, which has its origin in the resource-theory formulation of thermodynamics. Here, one considers a more general class of transformations known as thermal operations, corresponding to evolution in contact with a heat bath at temperature $T$. For example, one could envisage discharging the battery by using the stored mechanical energy to drive a current against a voltage bias, in which case the heat bath would comprise thermal electrons in the leads. The non-equilibrium free energy of the battery state is given by $\mathcal{F}_{\rm neq} = U(\hat{\rho}_b)-TS(\hat{\rho}_b)$, where $S(\hat{\rho}) = -\Tr [\hat{\rho}\ln \hat{\rho}]$ is the von Neumann entropy. This is generally greater than the free energy of the state at thermal equilibrium $\mathcal{F}_{\rm eq}=-k_{B}T\log{\mathcal{Z}}$ with $\mathcal{Z}=\Tr{\exp(-\hat{H}_b/k_BT)}$. The maximum extractable work under a thermal operation is given by the difference
\begin{equation}
    \label{W_free_energy}
    \mathcal{W}_F=\mathcal{F}_{\rm neq}-\mathcal{F}_{\rm eq},
\end{equation}
so that any non-thermal state is in principle a resource from which work may be extracted in this framework. Since thermal operations may be non-unitary, we have $\mathcal{W}_E\le \mathcal{W}_F$.

	\begin{figure}
		% Center the figure.
		\begin{center}
		% Include the eps file, scale it such that it's width equals the column width. You can also put width=8cm for example...
		\hspace*{-0.6cm}
		\includegraphics[width=\columnwidth]{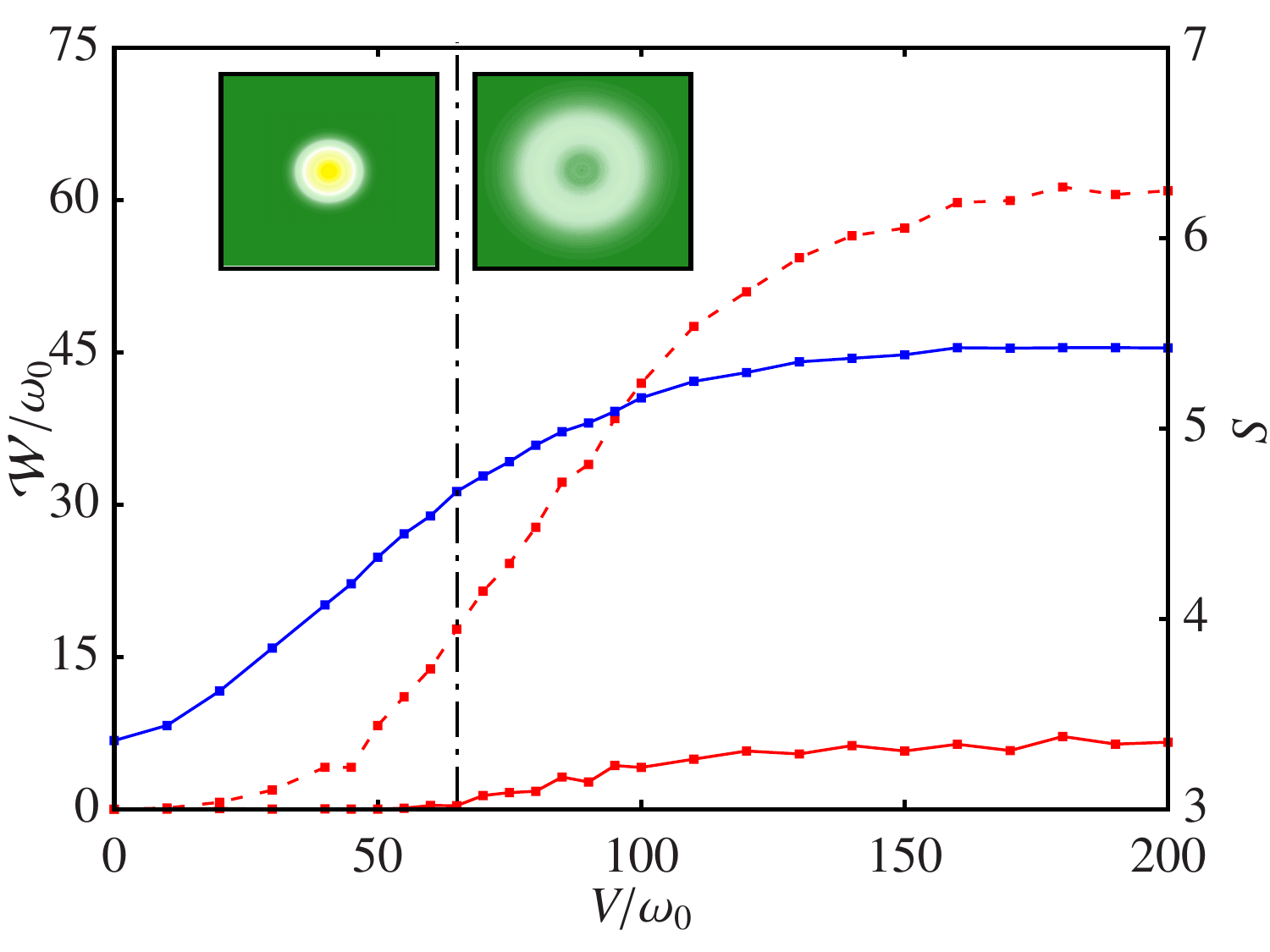}
		% Create a subtitle for the figure.
		\caption{Total work extractable from the steady state of the vibrational mode via a unitary transformation ($\mathcal{W}_E$, red, solid) and a thermal operation ($\mathcal{W}_F$, red, dashed). The threshold voltage is indicated by the vertical dot-dashed line and the von Neumann entropy of the oscillator is shown in blue. The same parameters as Fig.~\ref{fig:Wigner_function} are used.\label{Thermoplots}}
		% Define the label of the figure. It's good to use 'fig:title', so you know that the label belongs to a figure.
		\label{fig:tf_plot}
		\end{center}
	\end{figure}

In our case, the battery is a vibrational mode with Hamiltonian $\hat{H}_b = \omega_0 \hat{a}^\dagger\hat{a}$ and we consider the work extractable from its steady state $\hat{\rho}_b$. The results of this analysis are presented in Fig.~\ref{Thermoplots}. The energy of the oscillator increases monotonically with voltage, as expected. This is reflected in the behaviour of the the von Neumann entropy, which grows rapidly below threshold and eventually saturates far above threshold. We now turn to the ergotropy (red, solid) and the result is striking. Below threshold the ergotropy is exactly zero, meaning the non-thermal state of the oscillator is passive, while above threshold the state immediately becomes non-passive and $\mathcal{W}_E$ increases monotonically until saturation. Remarkably, therefore, we conclude that the ergotropy serves as an order parameter for the phonon lasing transition --- unlike, for example, the mode quadratures themselves --- so that the electromechanical battery only stores useful work above threshold. This is a key finding of our work. Meanwhile, the non-equilibrium free energy difference $\mathcal{W}_F$ is shown as red, dashed. This follows a similar trend to the ergotropy but with one exception: even the passive oscillator steady states below threshold may yield some work output under thermal operations due to their distance from thermal equilibrium. However, the value of $\mathcal{W}_F$ below threshold is relatively small, demonstrating the importance of self-sustained oscillations for effective battery charging.

\textit{Conclusions.---}We have quantified the conversion of electrical power into mechanical work stored in the quantum state of a NEMS device. Our results demonstrate that the self-sustained oscillations that were recently observed in Refs.~\cite{Wen2019,Urgell2019} can be interpreted as an electromechanical battery, thus bridging a gap between quantum thermodynamic theory and mesoscopic device physics.  We found that typical battery states in the quasi-adiabatic regime are diagonal in the energy eigenbasis but it may be possible to engineer configurations that generate mechanical coherence --- from which additional work can be extracted~\cite{Korzekwa2016,Francica2020} --- for example, by working in the strong electromechanical coupling regime~\cite{vigneau2021}. A pressing question for future work is to devise concrete protocols for transferring energy from the non-equilibrium battery states back into the electronic circuit~\cite{Lorch_2018}, opening up the exciting prospect of on-demand energy storage and extraction using mechanical degrees of freedom. Such electromechanical energy conversion could also be performed using heat rather than charge currents~\cite{Vikstrom2016}. Finally, we note that the coherent oscillations above threshold could be harnessed to implement an autonomous quantum clock~\cite{Erker_2017,Mitchison_2019, Milburn2020,Schwarzhans2021,Pearson2021}, whose thermodynamics will be analysed in a forthcoming publication.

\textit{Acknowledgements.---} We thank N.~Ares,  M.~Kewming, M.~Paternostro, M. Perarnau-Llobet, and A.~Purkayastha for useful discussions. This work was funded by the Irish Research Council, a Science Foundation Ireland-Royal Society University Research Fellowship, the European Research Council Starting Grant ODYSSEY (Grant Agreement No. 758403), and the EPSRC-SFI joint project QuamNESS. We acknowledge the Irish Centre for High End Computing for the provision of computational facilities. Some calculations were performed on the Kelvin cluster maintained by the Trinity Centre for High Performance Computing. This cluster was funded through grants from the Higher Education Authority, through its PRTLI program.

\bibliographystyle{apsrev4-1}
\bibliography{oisin.bib,bib.bib}

\clearpage 

\appendix
\section{\label{app:adiabatic_elim} Derivation of the Fokker-Planck equation}

In this Appendix we derive a Fokker-Planck equation for the Wigner function of the mechanical mode, which is equivalent to the Langevin equation~\eqref{langevin_equation} in the main text. Compressing all the electronic components of the Hamiltonian into a single term, $\hat{H}_\text{el} = \hat{H}_S + \hat{H}_B + \hat{H}_T$, we have
\begin{equation}
     \hat{H} = \frac{\hat{p}^2}{2m} + \frac{m\omega_0^2\hat{x}^2}{2} -F\hat{x}\hat{n} + \hat{H}_\text{el}.
\end{equation}
We introduce the Wigner transform of the density matrix,
\begin{equation}
\label{Wigner_Transform}
\r(x,p) = \frac{1}{\pi} \int \dd y \, \bra{x+y} \r \ket{x-y} \ee^{-\ii 2yp},
\end{equation}
which is an operator-valued function of $x$ and $p$ that acts on the electronic Hilbert space. The Wigner function of the oscillator follows from $W(x,p) = \Tr[\r(x,p)]$, while the electronic state is given by $\r_{\rm el} = \int\dd x\int\dd p \,\r(x,p)$. 

The time evolution of the system follows the Liouville-von Neumann equation $\text{d}\hat{\rho}/\text{d}t = -\ii[\hat{H},\rho]$. Using the operator correspondences for the Wigner function~\cite{GardinerZoller}, we obtain the corresponding equation of motion
\begin{align}
	\label{Wigner_Liouville_total}
	\pd{}{t}\r(x,p) & = -\ii \left [\hat{H}_{\rm el} - Fx \hat{n}, \r(x,p)\right ] \notag \\
		& \quad -\frac{p }{m} \pd{}{x} \r(x,p) + m\omega_0^2x \pd{}{p} \r(x,p)\notag \\
 	& \quad - \frac{F}{2} \pd{}{p} \left \{ \hat{n},\r(x,p)\right \} .
\end{align}
Physically, the first line represents the evolution of the electronic degrees of freedom conditioned on a particular oscillator position $x$, the second line represents the free mechanical evolution, and the third line represents the force on the oscillator due to the electronic charge. 

Now, our key assumption is that the electronic degrees of freedom, evolving under the Hamiltonian $\hat{H}_{\rm el} - Fx\hat{n}$, relax to a stationary state much faster than the characteristic evolution timescale of the oscillator. This is ensured by the timescale separation of Ineq.~\eqref{adiabatic_condition},
which justifies the adiabatic elimination of the electronic degrees of freedom from the mechanical equations of motion. Following the standard procedure~\cite{Gardiner}, the elimination is carried by defining the following operators:
\begin{align}
\label{L_el}
	\mathcal{L}_x & = -\ii [\hat{H}_{\rm el} - Fx \hat{n}, \bullet], \\
	\mathcal{H} & = \left ( m\omega_0^2 x - F \langle \hat{n}\rangle_x \right )  \pd{}{p} - \frac{p}{m} \pd{}{x}, \\
	\mathcal{V} & = -\frac{F}{2} \pd{}{p} \left \{ \hat{n} - \langle \hat{n}\rangle_x, \bullet \right \},
\end{align}
such that $\partial_t\hat{\rho}(x,p) = (\mathcal{L}_x+\mathcal{H} + \mathcal{V})\hat{\rho}(x,p)$. Here, $\langle \bullet\rangle_x = \Tr[\bullet \r_x]$ denotes an average with respect to the electronic stationary state, $\r_x$, defined by
\begin{equation}
\label{stationary_state}
\r_x = \lim_{t\to\infty} \ee^{\mathcal{L}_x t} \r_0 \r_B,
\end{equation}
where $\r_B$ denotes the initial state of the leads and $\r_0$ is an arbitrary initial reference state for the electronic system $S$, i.e.~the local charge degrees of freedom coupled to the oscillator. Here, we take the leads to be infinitely large and prepared in equilibrium at certain temperatures and chemical potentials, i.e.~$\r_B \propto \prod_{\alpha} \ee^{-\beta_\alpha(\hat{H}_\alpha - \mu_\alpha\hat{N}_\alpha)}$, where $\hat{H}_\alpha$ and $\hat{N}_\alpha$ are respectively the Hamiltonian and number operator of lead $\alpha$, while $\beta_\alpha$ and $\mu_\alpha$ are the corresponding inverse temperatures and chemical potentials. Strictly speaking, the steady-state density operator $\hat{\rho}_x$ describes the entire system-lead composite and exists only asymptotically (e.g.~in the sense of the McLennan-Zubarev ensemble~\cite{McLennan1959,Zubarev1970}). However, we will shortly see that the only relevant properties of $\r_x$ are those describing the vicinity of the localised charge (in particular, the noise spectrum associated with $\hat{n}$), which can be meaningfully assumed to relax quickly to the stationary state. Note that $\r_x$ depends on $x$ but not on $p$, and satisfies $\mathcal{L}_x\r_x = 0$ and $\Tr[\r_x] = 1$.

Let us now define a projector on the space of operator-valued functions of $x$ and $p$, as
\begin{equation}
\label{projector}
\mathcal{P} \hat{A}(x,p) = \Tr \left [\hat{A}(x,p)\right ] \r_x,
\end{equation}
whose orthogonal complement is denoted by $\mathcal{Q} = 1-\mathcal{P}$.  The projector obeys the relations
\begin{subequations}
	\label{projector_properties}
\begin{align}
	\label{projector_defining_property}
	\mathcal{P}^2 & = \mathcal{P}, \\
	\label{projector_kernel}
	\mathcal{P}\mathcal{L}_x & = 0 = \mathcal{L}_x \mathcal{P} , \\
	\label{projector_no_mean}
	\mathcal{P}\mathcal{V}\mathcal{P} & = 0,\\
	\label{Ham_proj_commutator}
	[\mathcal{P},\mathcal{H}] & = \frac{p}{m} \pd{\r_x}{x}\Tr[\bullet], \\
	\label{proj_comm_vanish}
	\mathcal{P}[\mathcal{P},\mathcal{H}] &= 0.
\end{align}
\end{subequations}
Eq.~\eqref{projector_defining_property} is the defining property of a projector, which follows here from the normalisation of $\r_x$. The first equality in Eq.~\eqref{projector_kernel} follows from the definition~\eqref{stationary_state} of $\r_x$, while the second equality holds since $\mathcal{L}_x$ generates a trace-preserving evolution. Eq.~\eqref{projector_no_mean} follows from the fact that the operator $\hat{n} - \langle \hat{n}\rangle_x$ has vanishing mean in the stationary state $\r_x$, which itself is independent of $p$. Finally, Eq.~\eqref{Ham_proj_commutator} can be proved by a direct calculation, while Eq.~\eqref{proj_comm_vanish} is satisfied because $\partial\r_x/\partial x$ is traceless. 

Let $\hat{r} = \mathcal{P}\r(x,p)$ denote the projected density matrix and $\hat{q} = \mathcal{Q}\r(x,p)$ its orthogonal complement. Applying these projectors to the left-hand side of Eq.~\eqref{Wigner_Liouville_total}, inserting appropriate factors of $1=\mathcal{P}+\mathcal{Q}$, and using the properties~\eqref{projector_properties}, we obtain
\begin{align}
	\label{P_eqn}
	\pd{\hat{r}}{t} & = \mathcal{P}\mathcal{H}\hat{r} +\mathcal{P}\left (\mathcal{H} + \mathcal{V}\right )\hat{q}, \\
	\label{Q_eqn}
	\pd{\hat{q}}{t} & = \mathcal{L}_x \hat{q} + \mathcal{Q}\left (\mathcal{H} + \mathcal{V}\right )\hat{q} + \mathcal{Q}(\mathcal{H} + \mathcal{V})\hat{r}.
\end{align}
Since the adiabatic condition implies that $\mathcal{L}_x \gg \mathcal{H},\mathcal{V}$, we see that $\hat{q}$ relaxes quickly to its steady state and can therefore be approximately replaced in Eq.~\eqref{P_eqn} by its stationary value. To proceed, we formally integrate Eq.~\eqref{Q_eqn} to obtain
\begin{align}\label{Q_soln}
& \hat{q}(t) = \mathcal{G}(t)\hat{q}(0) + \int_0^t\dd t' \mathcal{G}(t-t') \mathcal{Q}\left (\mathcal{H + \mathcal{V}}\right )\hat{r}(t'), \\
\label{G_def}
& \mathcal{G}(t) = \exp \left \lbrace\left [ \mathcal{L}_x + \mathcal{Q} (\mathcal{H}+\mathcal{V})\right]t \right \rbrace.
\end{align}
We now make the crucial assumption that $\hat{q}(0) = 0$, i.e. the electronic degrees of freedom begin in the stationary state $\r_x$ (or reach it after a negligibly short time). Substituting the solution~\eqref{Q_soln} into Eq.~\eqref{P_eqn} and shifting the integration variable $t'\to t-t'$ yields
\begin{align}
	\label{P_closed}
	\pd{\hat{r}}{t} & = \mathcal{P}\mathcal{H}\hat{r}(t) +\int_0^t\dd t' \mathcal{P}\left (\mathcal{H} + \mathcal{V}\right ) \mathcal{G}(t') \mathcal{Q}\left (\mathcal{H + \mathcal{V}}\right )\hat{r}(t-t')\notag\\  
	& \approx \mathcal{P}\mathcal{H}\hat{r}(t) + \int_0^\infty \dd t' \mathcal{P}\left (\mathcal{H} + \mathcal{V}\right ) \ee^{\mathcal{L}_x t' }\mathcal{Q}\left (\mathcal{H + \mathcal{V}}\right )\hat{r}(t).
\end{align}
On the second line, we have made the following approximations: (i)~since $ \mathcal{H},\mathcal{V} \ll \mathcal{L}_x$ we write $\mathcal{G}(t) \approx \ee^{\mathcal{L}_x t}$ to lowest non-trivial order in  small quantities (Born approximation), (ii)~assuming that the integrand decays rapidly as a function of $t'$, we approximate $\hat{r}(t-t') \approx \hat{r}(t)$ and extend the upper integration limit to infinity (Markov approximation).

To further simplify the integrand, we use Eqs.~\eqref{projector_properties} and the obvious property $\mathcal{P}\mathcal{Q} = 0$ to write
\begin{align}
&	\mathcal{P}\left (\mathcal{H} + \mathcal{V}\right )\ee^{\mathcal{L}_x t' }\mathcal{Q}(\mathcal{H} + \mathcal{V})\mathcal{P} \notag \\
& = \mathcal{P}\left ([\mathcal{P},\mathcal{H}] +\mathcal{H}\mathcal{P} + \mathcal{V}\right )\mathcal{Q}\ee^{\mathcal{L}_x t' } \mathcal{Q}(\mathcal{P}\mathcal{H} - [\mathcal{P},\mathcal{H}]+ \mathcal{V})\mathcal{P} \notag \\
& =  \mathcal{P} \mathcal{V}\ee^{\mathcal{L}_x t' } (\mathcal{V}-[\mathcal{P},\mathcal{H}])\mathcal{P}.
\end{align}
Tracing over the electronic degrees of freedom finally yields a formal master equation for the Wigner function, $W(x,p) = \Tr[\r(x,p)]$, in the form
\begin{equation}\label{master_equation_proj}
	\pd{}{t}W = \mathcal{H}W - \int_0^\infty\dd t'\, \Tr \left [ \mathcal{V} \ee^{\mathcal{L}_x t' } \left (\mathcal{V} - [\mathcal{P},\mathcal{H}]\right )\r_x\right ] W .
\end{equation}
The first term describes the free evolution of the oscillator while the second term describes diffusion and damping due to the fluctuating charge. 

It is straightforward to evaluate
\begin{align}
	\label{diffusion_evaluation}
\int_0^\infty \dd t\, \Tr \left [ \mathcal{V} \ee^{\mathcal{L}_x t}\mathcal{V} \r_x \right ] & = \frac{1}{2} D(x) \frac{\partial^2}{\partial p^2},
\end{align}
where the diffusion coefficient is given by 
\begin{equation}
\label{diffusion_coefficient}
D(x) = F^2 \int_0^\infty \dd t\, \braket{\left \{ \delta\hat{n}(t), \delta\hat{n}(0)\right \}}_x,
\end{equation}
and we defined $\delta \hat{n}(t) = \hat{n}(t) - \braket{\hat{n}}_x$ as the charge fluctuation operator evaluated in the Heisenberg picture, with $\hat{n}(t) = \ee^{\mathcal{L}_x^\dagger t} \hat{n}$ and $\mathcal{L}_x^\dagger = \ii [\hat{H}_{\rm el} -F x \hat{n}, \bullet]$ the adjoint Liouvillian. The remaining contribution to Eq.~\eqref{master_equation_proj} is evaluated as
\begin{equation}\label{damping_evaluation}
	\int_0^\infty \dd t\, \Tr \left [ \mathcal{V} \ee^{\mathcal{L}_x t}[\mathcal{P},\mathcal{H}] \r_x\right ] = \gamma(x) \pd{}{p} p,
\end{equation}
where we defined the damping rate
\begin{equation}\label{damping_rate}
\gamma(x) = \frac{F}{m} \int_0^\infty\dd t\, \Tr \left [ \hat{n}(t) \pd{\r_x}{x}\right ].
\end{equation}

We express $\gamma(x)$ in terms of the charge noise spectrum as follows. First, note that consistency with the Markov approximation requires the integrand of Eq.~\eqref{damping_rate} to decay to zero for large times. Any differentiable function $g(t)$ that vanishes at infinity, $\lim_{t\to \infty} g(t)=  0$, satisfies
\begin{align}
	\label{integral_identity}
&	-\int_0^\infty \dd t\, g(t) = \int_0^\infty\dd t \int_t^\infty\dd t' \, \od{g}{t'}  =\int_0^\infty\dd t\,  t \od{g}{t},
\end{align}
where the first equality follows from the definition of integration and the second equality follows after switching the order of integration, $\int_0^\infty\dd t\, \int_{t}^\infty \dd t'  = \int_0^\infty\dd t'\, \int_0^{t'}\dd t$, then carrying out the trivial integral over $t$ and relabelling dummy variables. For the particular case of $g(t) = \Tr[\hat{n}(t) \partial\r_x/\partial x]$, we have $\dd g/\dd t = \Tr[\hat{n}(t) \mathcal{L}_x \partial\r_x/\partial x]$. By differentiating $\mathcal{L}_x\r_x$ with respect to $x$, we obtain
\begin{equation}\label{deriv_identity}
	\mathcal{L}_x \pd{ \r_x}{x}  = -\ii F[\hat{n}, \r_x].	
\end{equation}
Putting everything together, we get 
\begin{equation}\label{damping_rate_correlation}
\gamma(x) = \frac{F^2}{m} \int_0^\infty\dd t\, \braket{\ii t \left [\hat{n}(t),\hat{n}(0)\right ]}_x.
\end{equation} 
Eqs.~\eqref{diffusion_coefficient} and~\eqref{damping_rate_correlation} are equivalent to Eq.~\eqref{diffusion_damping} in the main text.

It is now straightforward to write down the equation of motion for the Wigner function
\begin{align}\label{Fokker_Planck}
	\pd{W}{t} & = \pd{}{p}  \left (m \omega_0^2 x - F\braket{\hat{n}}_x + \gamma(x) p  \right )W  - \frac{p}{m} \pd{W}{x} \notag \\ 
	& \quad + \frac{D(x)}{2}  \frac{\partial^2W}{\partial p^2} .
\end{align}
Eq.~\eqref{Fokker_Planck} is a genuine Fokker-Planck equation for the Wigner function of the oscillator. Its solution may take negative values if the initial condition for $W(x,p)$ is non-classical. However, whenever $W(x,p)\geq 0$ it may be interpreted as a probability distribution associated with the Langevin equation~\eqref{langevin_equation}; see Ref.~\cite{Gardiner}. The Wigner function is then reconstructed as
\begin{equation}
    W(x',p,t) = \mathbb{E}[\delta(x'-x(t))\delta (p-m\dot{x}(t))], 
\end{equation}
where $\mathbb{E}[\bullet]$ denotes an average over Langevin trajectories $x(t)$ evolved up to a given time $t$. The Wigner function generates expectation values with symmetric $\hat{x}-\hat{p}$ operator ordering~\cite{GardinerZoller}, such as 
\begin{align}
    \label{Wigner_moments}
    &\Tr [\hat{x}\hat{\rho}(t)] = \int\dd x\int \dd p \, x W(x,p,t),\\ 
    &\Tr [\hat{p}\hat{\rho}(t)] = \int\dd x\int \dd p \, p W(x,p,t).
\end{align}

The Langevin equation can also be used to reconstruct symmetrised two-time correlation functions, such as $C(t,t+\tau) = \tfrac{1}{2}\langle \{\hat{x}(t+\tau),\hat{x}(t)\}\rangle = \Re \langle \hat{x}(t+\tau)\hat{x}(t)\rangle$. Using the Born-Markov assumption and the quantum regression formula~\cite{GardinerZoller}, this can be written as 
\begin{equation}
    \label{regrssion_formula}
    \langle \hat{x}(t+\tau)\hat{x}(t)\rangle = \Tr\left [ \hat{x} \mathcal{E}(t+\tau,t) \left(\hat{x} \hat{\rho}(t) \right)\right],
\end{equation}
where $\mathcal{E}(t+\tau,t)$ is the time evolution superoperator. In phase space, the expression $\mathcal{E}(t+\tau,t) \left(\hat{x} \hat{\rho}(t) \right)$ corresponds to the solution of the Fokker-Planck equation~\eqref{Fokker_Planck} with initial condition
\begin{equation}
\label{initial_cond_xrho}
\hat{x}\hat{\rho}(t) \to \bar{W}(x,p,t) = x W(x,p) + \frac{\ii}{2}\frac{\partial W(x,p,t)}{\partial p},
\end{equation}
where $W(x,p,t)$ is the Wigner function associated with $\hat{\rho}(t)$. This evolves to
\begin{equation}
\label{evolved_xrho}
\int\dd x'\dd p' \Pi(x,p,t+\tau|x',p',t)\bar{W}(x',p',t),
\end{equation}
with $\Pi(x,p,t+\tau|x',p',t)$ the Green function of the Fokker-Planck equation, i.e.~its formal solution for a point initial condition $\delta(x-x')\delta(p-p')$. Taking the real part of Eq.~\eqref{regrssion_formula} and using Eq.~\eqref{Wigner_moments}, we thus obtain
\begin{widetext}
\begin{align}
  \Re \langle \hat{x}(t+\tau)\hat{x}(t)\rangle & = \int\dd x\int\dd x'\int\dd p\int \dd p'\, x  \Pi(x,p,t+\tau|x',p',t) x' W(x',p',t) \notag \\
  & = \int\dd x\int\dd x'\, x x'  P(x,t+\tau;x',t)\notag \\
  & = \mathbb{E}[x(t+\tau)x(t)],
\end{align}
\end{widetext}
where \begin{equation} 
P(x,t+\tau;x',t) = \int \dd p\!\int \dd p'\, \Pi(x,p,t+\tau|x',p',t) W(x',p',t)
\end{equation}
is the joint probability distribution for the random variables $x(t+\tau)$ and $x(t)$ given that $x(t)$ is distributed according to $ P(x,t) = \int\dd p\,W(x,p,t)$.

Finally, let us briefly recap the assumptions underlying Eq.~\eqref{Fokker_Planck}. We assume that the electronic degrees of freedom are initially in the stationary state, $\mathcal{Q}\hat{\rho}(0) = 0$, and relax quickly back to this state when perturbed, $\mathcal{L}_x\gg \mathcal{H},\mathcal{V}$. These assumptions are well satisfied in the quasi-adiabatic regime defined by Eq.~\eqref{adiabatic_condition}. Eq.~\eqref{Fokker_Planck} then follows from a Born-Markov approximation for the projected density matrix~\eqref{projector}. Note, however, that we make no assumptions about the Hamiltonian $\hat{H}_{\rm el}$ of the electronic system or the operator $\hat{n}$, which could in principle describe an arbitrary interacting system so long as the quasi-adiabatic assumption is satisfied.
 
\section{\label{app:NEGF} Solving for the electronic steady state}

To solve the Langevin equation one needs to calculate the occupation $\langle \hat{n}\rangle_x$ and the noise spectrum $S_x(\omega)$ for the central system in its steady state. To achieve this we use an equation of motion approach.

The electronic Hamiltonian conditioned on the oscillator position $x$ is defined by
\begin{equation}
    \label{H_elec}
    \hat{H}_x = \hat{H}_S + \hat{H}_B + \hat{H}_T - Fx\hat{c}^\dagger\hat{c},
\end{equation}
where $\hat{H}_S + \hat{H}_B + \hat{H}_T$ is given in Eqs.~\eqref{ham_quantum_dot}--\eqref{ham_tunnelling}. We ignore the constant term proportional to $N_0$ since this merely shifts the zero of energy. Eq.~\eqref{H_elec} describes a resonant-level model comprising a quantum dot coupled to two electrodes, whose analysis can be found in many texts, e.g.~Refs.~\cite{ryndyk2016,Schaller_2014}. The Heisenberg equations of motion for this system read
\begin{align}
\label{c_eom}
    & \frac{\text{d}}{\text{d}t}\hat{c} = -\ii\left(\epsilon_x\hat{c} + \sum_{\alpha\in L,R}\sum_k g_{k\alpha}\hat{d}_{k\alpha} \right), \\
    \label{d_eom}
    & \frac{\text{d}}{\text{d}t}\hat{d}_{k\alpha} = -\ii\left(\Omega_{k\alpha}\hat{d}_{k\alpha} + g_{k\alpha}\hat{c}\right),   
\end{align}
where $\epsilon_x = \epsilon - F x$. Formally integrating Eq.~\eqref{d_eom} and inserting the result into Eq.~\eqref{c_eom}, one finds
\begin{equation}
\label{quantum_langevin_equation}
    \frac{\text{d}\hat{c}}{\text{d}t} = -\ii\epsilon_x\hat{c}(t) + \sum_\alpha\left( \hat{\zeta}_\alpha(t) - \int^\infty_{t_0}\dd t'\chi_\alpha(t-t')\hat{c}(t') \right),
\end{equation}
where we introduced the memory kernel $\chi_\alpha(t)$ and the noise operator $\hat{\zeta}_\alpha(t)$, defined by
\begin{align}
    &\chi_\alpha(t-t') = \Theta(t-t')\sum_k g^2_{k\alpha}e^{-\ii\Omega_{k\alpha}(t-t')}\\
    &\hat{\zeta}_\alpha(t) = -\ii\sum_kg_{k\alpha}e^{-\ii\Omega_{k\alpha}(t-t_0)}\hat{d}_{k\alpha}(t_0),
\end{align}
with $t_0$ defining the initial time at which the leads are prepared in thermal equilibrium. Since we are interested in the steady state, we take the limit $t_0\to -\infty$. Eq.~\eqref{quantum_langevin_equation} is then easily solved in the Fourier domain to obtain
\begin{equation}
\label{Fourier_solution}
    \tilde{c}(\omega) = G(\omega)\tilde{\zeta}(\omega),
\end{equation}
where $G(\omega)$ is the retarded Green function
\begin{equation}
\label{Green_function}
    G(\omega) = [\omega - \epsilon_x - \tilde{\chi}_L(
    \omega) - \tilde{\chi}_R(\omega)]^{-1},
\end{equation}
with $\tilde{\chi}_\alpha$ the self energy of electrode $\alpha = L,R$,
\begin{equation}
    \tilde{\chi}_\alpha(\omega) = \fint \frac{\text{d}\omega'}{2\pi} \frac{\kappa_\alpha(\omega')}{\omega-\omega'} - \ii \frac{\kappa_\alpha(\omega)}{2},
\end{equation}
where $\fint$ is a principal-value integral and $\kappa_\alpha$ is the spectral density of the quantum dot coupled with the left or right electrode.

The non-equilibrium steady state depends on the statistical properties of the noise operator. The initial state of the reservoirs is assumed to be thermal and uncorrelated, i.e.~$\langle\hat{d}_{\alpha,k}^\dagger(t_0)\hat{d}_{\alpha',k'}(t_0)\rangle=f_\alpha(\Omega_k)\delta_{\alpha\alpha'}\delta_{kk'}$, where $f_\alpha(\omega) = (\ee^{\beta_\alpha(\omega-\mu_\alpha)}+1)^{-1}$ is the Fermi-Dirac distribution. The statistics of the noise operator can be found in the frequency domain, where
\begin{equation}
    \tilde{\zeta}_\alpha(\omega) = -\sqrt{2\pi}i\sum_kg_{k\alpha}\delta(\omega-\Omega_{k\alpha})\hat{d}_{k\alpha}(t_0).
\end{equation}
One can calculate the two-point noise spectrum to be 
\begin{equation}
    \langle \tilde{\zeta}^\dagger_\alpha(\omega) \tilde{\zeta}_{\alpha'}(\omega') \rangle = \kappa_\alpha(\omega)\delta(\omega-\omega')f_\alpha(\omega)\delta_{\alpha\alpha'}
\end{equation}
Substituting this back into Eq.~\eqref{Fourier_solution} of the dot 
\begin{equation}
    \langle \tilde{c}^\dagger(\omega)\tilde{c}(\omega')\rangle = \sum_{\alpha \in L,R} A(\omega)\kappa_\alpha(\omega)f(\omega),
\end{equation}
where $A(\omega) = |G(\omega)|^2$ is the spectral function. To find the steady state occupation we evaluate the Fourier transform at the time $t=t'$ to obtain
\begin{equation}
    \langle \hat{c}^\dagger \hat{c}\rangle_x = \frac{1}{2\pi}\sum_{\alpha\in L,R} \int A(\omega)\kappa_\alpha(\omega)f_\alpha(\omega).
\end{equation}
For the parameters we consider, with $\tfrac{1}{2}(\mu_L+\mu_R) = \epsilon$ so that the dot is half-filled on average, this yields $N_0 =  \langle \hat{c}^\dagger \hat{c}\rangle_0=\tfrac{1}{2}$ and thus $\langle \hat{n}\rangle_x = \langle \hat{c}^\dagger \hat{c}\rangle_x-\tfrac{1}{2}$.

To find the dissipation rate and diffusion coefficient one needs to calculate the charge noise spectrum
\begin{equation}
\label{noise_func}
S_x(\omega) = F^2\int^\infty_{-\infty}\text{d}t\, e^{i\omega t} \left[\langle \hat{n}(t) \hat{n}(0)\rangle_x - \langle \hat{n}\rangle_x^2\right].
\end{equation}
One thus needs to calculate the two-point correlation function of the operator $\hat{c}^\dagger\hat{c}$:
\begin{equation}
    \langle\hat{c}^\dagger(t)\hat{c}(t)\hat{c}^\dagger(0)\hat{c}(0)\rangle
\end{equation}
We proceed using a similar method to the occupation function using a Green's Function approach
\begin{align}
    \langle\hat{c}^\dagger(\omega)\hat{c}(\omega')\hat{c}^\dagger(\upsilon)\hat{c}(\upsilon')\rangle = & G^*(\omega)G(\omega')G^*(\upsilon)G(\upsilon') \nonumber\\
    &\langle  \hat{\zeta}^\dagger(\omega)\hat{\zeta}(\omega')\hat{\zeta}^\dagger(\upsilon)\hat{\zeta}(\upsilon')\rangle
\end{align}
To calculate $\langle  \hat{\zeta}^\dagger(\omega)\hat{\zeta}(\omega')\hat{\zeta}^\dagger(\omega'')\hat{\zeta}(\omega''')\rangle$ we make use of Wick's theorem using the identity 
\begin{align}
\label{Wick_theorem}
    \langle\hat{d}^\dagger_{k\alpha}\hat{d}_{k'\alpha'}\hat{d}^\dagger_{j\beta}\hat{d}_{j'\beta'}\rangle = & \delta_{k'k}\delta_{\alpha'\alpha}f(\Omega_k)\delta_{j'j}\delta_{\beta'\beta}f(\Omega_j) +\nonumber\\
    &\delta_{kj'}\delta_{\alpha\beta'}f(\Omega_k)\delta_{k'j}\delta_{\alpha'\beta}\left[ 1-f(\Omega_j)\right]
\end{align}
The first term in this expression is equivalent to the product of the means and will cancel with the contribution from $\langle \hat{n}\rangle_x^2$ in Eq.~\eqref{noise_func}. The second term in equation~\eqref{Wick_theorem} is unique to the four-point correlation function, and its contribution to $\langle\hat{\zeta}^\dagger(\omega)\hat{\zeta}(\omega')\hat{\zeta}^\dagger(\upsilon)\hat{\zeta}(\upsilon')\rangle$ is
\begin{align}
  4\pi^2\sum_k\sum_j& |g_{k\alpha}|^2|g_{j\beta}|^2f_\alpha(\Omega_k)\left[1 - f_\beta(\Omega_j)\right] \nonumber\\
    \times&\,\delta(\omega+\upsilon')\delta(\omega'+\upsilon)\delta(\omega-\Omega_{k\alpha})\delta(\omega'-\Omega_{j\beta})
\end{align}
Using the same method as the two point correlation function, the contribution to $\langle\hat{c}^\dagger(\omega)\hat{c}(\omega')\hat{c}^\dagger(\upsilon)\hat{c}(\upsilon')\rangle$ is therefore
\begin{align}
    A(\omega)A(\omega')\kappa_\alpha(\omega)\kappa_\beta(\upsilon)f_\alpha(\omega)[1-f_\beta(\upsilon) ]\delta(\omega+\upsilon')\delta(\omega'+\upsilon).
\end{align}
Performing a Fourier transform, change of variables and letting $s=s'=0$ and $t=t'$ one arrives at the result
\begin{align}
    \langle\hat{n}(t)\hat{n}(0)\rangle_x - \langle\hat{n}\rangle_x^2 & = \frac{1}{4\pi^2}\int^\infty_{-\infty}\int^\infty_{-\infty}\text{d}\omega\text{d}\omega'\, \ee^{-\ii\omega t}A(\omega+\omega')A(\omega')\nonumber\\
    & \times \kappa_\alpha(\omega+\omega')\kappa_\beta(\omega')f_\alpha(\omega+\omega')\left[1-f_\beta(\omega') \right].
\end{align}
Inserting this into Eq.~\eqref{noise_func} one finds
\begin{align}
    S_x(\omega) =& F^2\sum_{\alpha,\beta}\int^\infty_{-\infty}\frac{\text{d}\omega'}{2\pi}A(\omega+\omega')A(\omega')\kappa_\alpha(\omega+\omega')\kappa_\beta(\omega')\nonumber\\
    &f_\alpha(\omega+\omega')[1-f_\beta(\omega')]
\end{align}
From Eq.~\eqref{diffusion_damping} in the main text, the diffusion coefficient is therefore
\begin{equation}
 D(x) = F^2\!\!\sum_{\alpha,\beta\in L,R}\int\frac{\text{d}\omega'}{2\pi}A(\omega')^2\kappa_\alpha(\omega')\kappa_\beta(\omega')f_\alpha(\omega')[1-f_\beta(\omega')],
\end{equation}
while the dissipation rate for the system is
\begin{align}
    m\gamma(x) = &F^2\sum_{\alpha,\beta} \int \frac{\dd\omega}{2\pi} A(\omega)\kappa_\alpha(\omega)f_\alpha(\omega) \nonumber\\
    \times\,&\frac{\text{d}}{\text{d}\omega'}\left(\kappa_\beta(\omega+\omega')A(\omega+\omega')\left[1-f_\beta(\omega+\omega')\right]\right)\Bigg\rvert_{\omega'=0}
\end{align}
One can thus perform these calculations for any system given the spectral density $\kappa_\alpha(\omega)$. For the Lorentzian spectral density~\eqref{Lorentzian_spectral_density} the electrode self energy is given by
\begin{equation}
    \tilde{\chi}_\alpha(\omega) = \frac{\Gamma_\alpha\delta_\alpha(\omega-\omega_{0\alpha})}{2\left[(\omega -\omega_{0\alpha})^2 + \delta_\alpha^2 \right]} - i\frac{\Gamma_\alpha\delta_\alpha^2}{2\left[(\omega-\omega_{0\alpha})^2+\delta_\alpha^2 \right]}.
\end{equation}
This is used in~\eqref{Green_function} to calculate the spectral function of the system.

% % Wigner Appendix

\section{\label{app:langevin} Numerical Simulation of the Langevin Equation}

The Langevin equation, equation 7 in text, was solved by using the stochastic Euler method. The space for the occupation, fluctuation and dissipation graphs were calculated using the formulae derived in appendix B. From these graphs it is possible to solve the Langevin equation using a stochastic Euler method\cite{Gardiner2004,Jabcobs2010}. A time-step $\Delta t$ was selected, the system was evolved to a time $t+\Delta t$ using the values of the oscillator at the position at time t. The evolution step from time $t$ to time $t + \Delta t$ is given by 
\begin{align}
    v(t+\Delta t) & = v(t) + \left(-\omega_0^2x(t) - \gamma[x(t)]v(t) + \langle n \rangle[x(t)] \right) \Delta t \nonumber\\
    & \quad + \sqrt{D[x(t)]}\Delta W,\\
    x(t+\Delta t) & = x(t) + v(t)\Delta t,
\end{align}
where $\Delta W$ is a Gaussian variable with mean 0 and variance $\Delta t$. Once an initial state of the oscillator is selected, an array of velocity and position values can be generated. From this array we can construct the Wigner Functions seen in Figure~\ref{fig:Wigner_function} of the main text. 

For each bias $2.5*10^8$ time steps were used. Here we assume that the system is ergodic - that we can sample the steady-state probability density via a single trajectory calculation and thus only perform one iteration of each calculation.

\section{\label{app:Wigner} The Wigner Function}

The Wigner function used throughout the paper represents a quasi-probability distribution similar to phase space in classical mechanics. Phase space explores ensembles of trajectories, while the Wigner space explores quantum probabilities. However, Wigner quasi-probability distributions can take negative values setting them apart from their classical counterparts. The Wigner function is given as the Fourier Transform of the symmetric characteristic function\cite{Barnett1997}
\begin{equation}
    W(\alpha) = \frac{1}{\pi^2} \int^\infty_{-\infty} d^2\xi \chi(\xi)\text{exp}(\alpha\xi^*-\alpha^*\xi),
\end{equation}
where $\alpha$ is a complex number with real position and imaginary momentum of the oscillator. The symmetric characteristic function ($\chi$), given by 
\begin{equation}
    \chi(\xi) = \text{Tr}\left[\hat{\rho}\hat{D}(\xi) \right]
\end{equation}
where $\hat{D}(x) = \text{exp}(x\hat{a}^{\dagger} -x^*\hat{a})$ is the Glauber displacement operator. The Wigner function acts as the closest parallel to the outcome of measurements of the $\hat{p}$ and $\hat{q}$ operator, thus one can set the initial state of the system and evolve the system to steady-state using the probability distribution as the Wigner Function of the system.\\

Once we have the steady state Wigner function, it is important to reconstruct the state back in density matrix formalism to evaluate the thermodynamics of the system. The calculations to convert the Wigner Function to the density matrix can be found in reference \cite{Paternostro2009}. The formula for each element of the the density matrix $\rho = \gamma_{n,m} \ket{n}\bra{m}$ is given by 
\begin{equation}
    \gamma_{n,m} = \frac{1}{\pi} \int d^2\zeta \chi(\zeta)f_{nm}(\zeta),
\end{equation}
where $\chi$ is the characteristic function of the system and $f_{nm}(\zeta) = \bra{n}\hat{D}(-\zeta)\ket{m}$. When n>m this is given by
\begin{equation}
    f_{nm} = \sqrt{\frac{m!}{n!}}(-\zeta)^{n-m}e^{-|\zeta|^2/2}\mathcal{L}^{n-m}_m(|\zeta|^2),
\end{equation}
where $\mathcal{L}^{n-m}_m(|\zeta|^2)$ is the generalised Laguerre polynomial of degree m and argument $|\zeta|^2$. For the case when m=n this simplifies to 
\begin{equation}
    f_{nn} = e^{-|\zeta|^2/2}\mathcal{L}^{0}_n(|\zeta|^2).
\end{equation}
To convert the Wigner Function to the Characteristic function one needs to perform an inverse Fourier transform
\begin{equation}
\chi(\zeta) = \frac{1}{\pi^2}\int \text{d}\alpha W(\alpha)\text{exp}(-\alpha \zeta* + \alpha*\zeta) .
\end{equation}
For all the systems in consideration in this paper the Wigner function appears to be radially symmetric, this simplifies the calculations for the density matrix in two ways: The original integral for the Wigner Function can be calculated in polar coordinates with the angular component able to be evaluated analytically. The set of equations then simplifies down to 
\begin{equation}
    \chi(r) = 2\pi \int du u \mathcal{J}_0(2ru)W(u),
\end{equation}
where u is the radial value for the second integral, $\mathcal{J}_0$ is the Bessel function of the first kind and r is the radial value of the generated Wigner function. 

The second simplification is that the off diagonal components for the density matrix are all 0, ie there are no coherences in the energy eigenbasis. This is due to the rotational symmetry of the system inferring that the density matrix of the system is not evolving over time. Because of these simplifying factors the density matrix could be calculated very quickly for a large number of terms for the system. The full equation describing the transform is therefore given by 
\begin{equation}
    \rho_{nn} = 8\pi \int\int dudr u \mathcal{J_0}(2ru)W(u) r e^{-|r|^2/2}\mathcal{L}^{0}_n(r^2).
\end{equation}

\section{\label{app:app} Fluctuation, Dissipation and Occupation Functions}

\begin{figure*}

\hspace*{-1.5cm}
\includegraphics[scale=0.9]{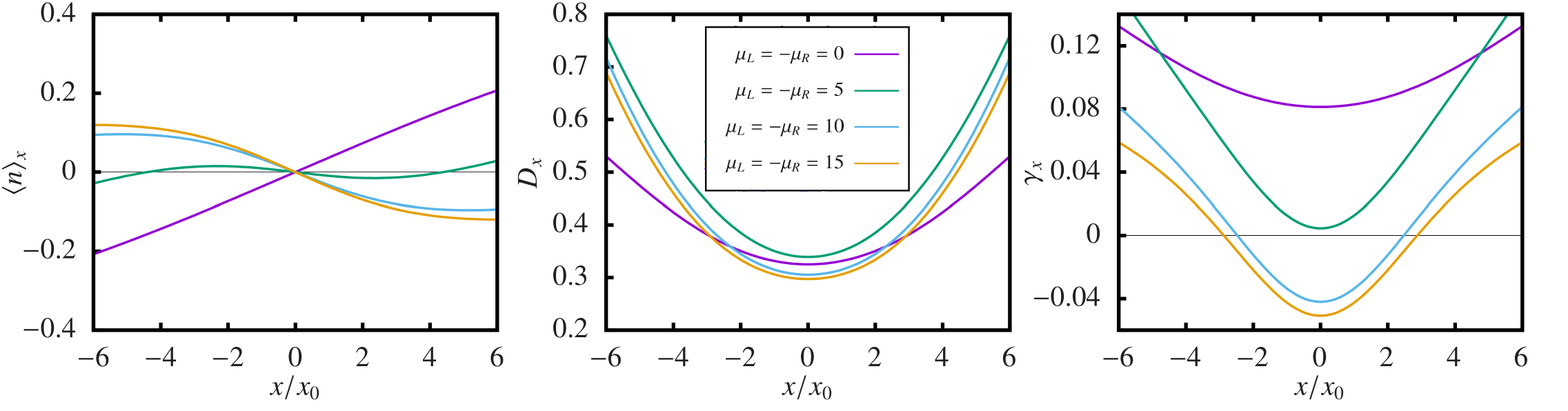}\par

\caption{Excess occupation, fluctuation and dissipation curves for various voltages as a function of the oscillator's position. a) The quantum dot's excess occupation. At zero voltage the occupation curve is monotonous with the oscillator position, as the voltage across the quantum dot is increased the curve becomes non-monotonic. b) The fluctuation curve at various voltages. c) The dissipation curve at various voltages. When the Occupation curve is monotonic, no negative damping occurs however once the voltage reaches a threshold and the occupation becomes nonmonotonic, the damping coefficient becomes negative. Parameters: $\omega_0=0.2$,$m=1$,$\lambda=0.1$,$\omega_L=0.5$,$\omega_R=-0.5$,$\delta_L=\delta_R=1$,$\Gamma_L=\Gamma_R=2$,$\beta_L=\beta_R=0.5$.\label{occ_diss_fluc}}
\end{figure*}

The fluctuation, dissipation and occupations functions were numerically calculated using the formulae outlined in appendix~\ref{app:NEGF}. As shown in Fig.~\ref{fluc_diss_theorem} at equilibrium the ratio between the fluctuation and dissipation is 4 for all positions, this implies the fluctuation dissipation theorem (FDT), $D(x)= 2m\gamma(x)/\beta$, is satisfied. This highlights that at equilibrium $\gamma(x)>0$. The FDT arises due to the detailed balance condition, $S_x(-\omega) = \ee^{-\beta\omega}S_x(+\omega)$. Out of equilibrium we require an energy dependent spectral density to achieve negative damping as outlined in \cite{Bennett2006}. We achieve this using a Lorentzian spectral density giving rise to the occupation, fluctuation and dissipation curves seen in Fig.~\ref{occ_diss_fluc}. Once a threshold voltage is reached, a region of the dissipation curve becomes negative leading to the lasing state of the system. 

\begin{figure*}

\hspace*{-1.5cm}
\includegraphics[scale=0.9]{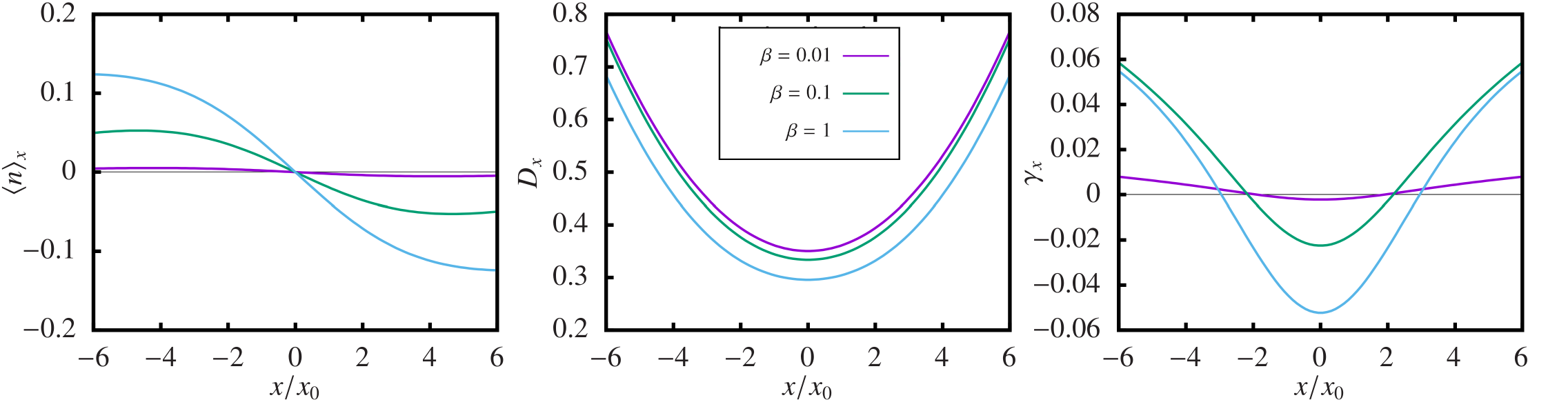}\par

\caption{Excess occupation, fluctuation and dissipation curves for various inverse temperatures as a Function of the Oscillator's position. a) The quantum dot's occupation. As the temperature is increased ($\beta$ is decreased) the excess occupation gets closer to 0 for all values of $x/x_0$. b) The fluctuation curve at various voltages, increases in temperature increases the fluctuations. c) The dissipation curve. The effect of higher temperature is to significantly weaken the strength of the negative dissipation as well as reducing the region of negative damping. Parameters: $\omega_0=0.2$,$m=1$,$\lambda=0.1$,$\omega_L=0.5$,$\omega_R=-0.5$,$\delta_L=\delta_R=1$,$\Gamma_L=\Gamma_R=2$,$\mu_L=-\mu_R=15$.\label{occ_diss_fluc_temp}}
\end{figure*}

The effect of temperature is shown in Fig.~\ref{occ_diss_fluc_temp}. As the temperature of the leads are increased ($\beta$ is decreased), the negative dissipation is mollified. As the temperature is increased the Fermi distributions of the leads become increasingly similar, the effect of this is to cause the excess charge on the oscillator to be closer to 0 for all values of the position as well as significantly weakening the strength of the negative dissipation as well as reducing the region of negative damping.

\begin{figure*}

\hspace*{-0.9cm}
\includegraphics[scale=0.7]{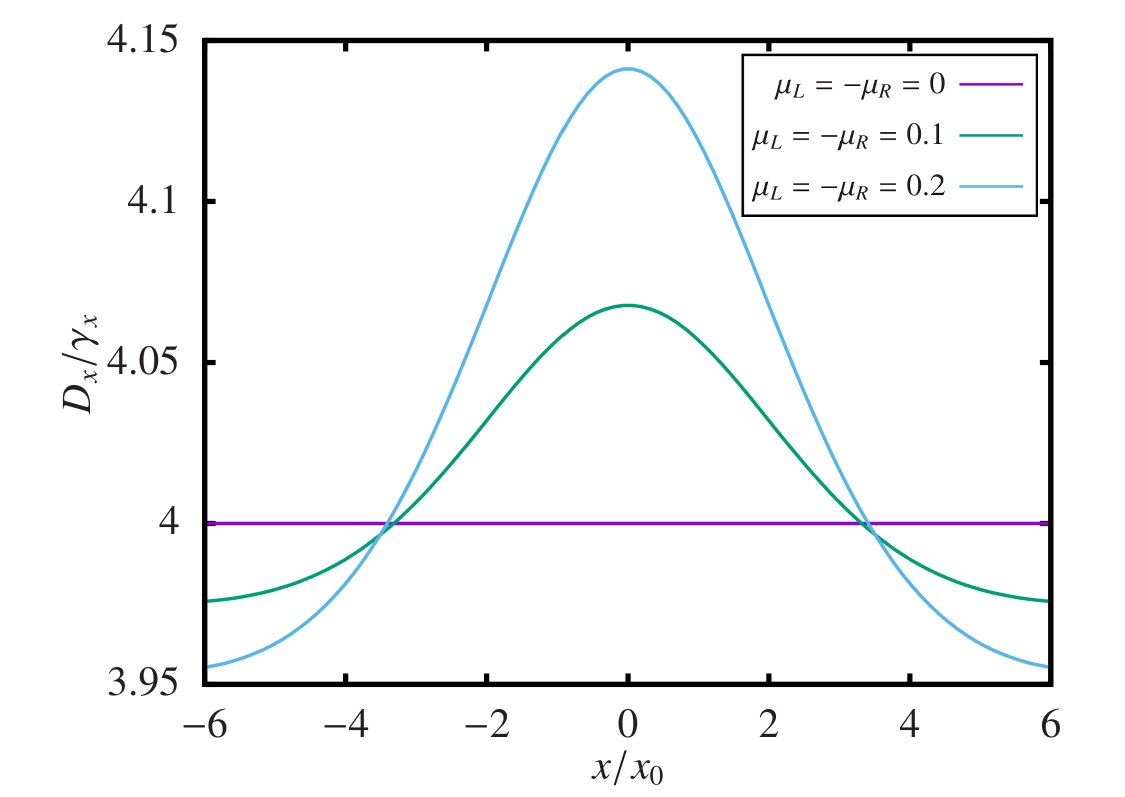}\par

\caption{The fluctuation dissipation ratio at various biases. When there is no applied bias, the Fluctuation Dissipation Theorem (FDT) - $D(x) =2mk_BT\gamma(x)$ - holds as shown. When the system is pushed away from equilibrium the FDT is no longer valid.\label{fluc_diss_theorem}}
\end{figure*}

%
%
%%%%%%%%%%%%%%%%%%%%%%%%%%%%%%%%%%%%%%%%%%%%%%%%%%%%%%%%%%%%%%%%%%%%%%%%

\end{document}